# Limiting efficiencies of solar energy conversion and photo-detection via internal emission of hot electrons and hot holes in gold

Svetlana V. Boriskina*, Jiawei Zhou, Wei-Chun Hsu, Bolin Liao, Gang Chen
Department of Mechanical Engineering, Massachusetts Institute of Technology
Cambridge, MA 02139, USA

## ABSTRACT

We evaluate the limiting efficiency of full and partial solar spectrum harvesting via the process of internal photoemission in Au-semiconductor Schottky junctions. Our results based on the ab initio calculations of the electron density of states (e-DOS) reveal that the limiting efficiency of the full-spectrum Au converter based on hot electron injection is below 4%. This value is even lower than previously established limit based on the parabolic approximation of the Au electron energy bands. However, we predict limiting efficiency exceeding 10% for the hot holes collection through the Schottky junction between Au and p-type semiconductor. Furthermore, we demonstrate that such converters have more potential if used as a part of the hybrid system for harvesting high- and low-energy photons of the solar spectrum.

**Keywords:** Hot carrier collection, Schottky junction, solar energy harvesting, internal photoemission, gold (Au)

## 1. INTRODUCTION

Harvesting solar energy by photon absorption in metal nanostructures followed by collection of photo-generated hot electrons via the processes of internal photoemission (IPE) has been recently explored as a promising alternative approach to traditional photovoltaics as well as for catalysis and photo-detection[1–7]. Traditionally, noble metals such as Au are considered as good candidates for gapless photon absorbers, which are potentially capable of full spectrum harvesting[5,8–10]. The principle of operation of the IPE energy converter is schematically illustrated in Fig. 1. Energy of absorbed photons raises the energy of electrons above the Fermi level, creating so-called hot electrons (Fig. 1a). Photo-generated hot electrons typically cool down very fast due to scattering on phonons, lattice defects, and cold electrons. The cooling process occurs on picosecond timescale in most metals. If, however, these hot electrons can be extracted before they cool down, they can contribute to the energy conversion efficiency.

Typically, such full-spectrum converters use the Schottky junction that forms at the interface between metal and n-type semiconductor as a frequency-selective filter to extract photo-generated hot electrons (Fig. 1a). As a result, hot electrons generated by absorption of photons with energies below the semiconductor bandgap can still be harvested by using this approach. This offers the way to potentially increase the conversion efficiency of photovoltaic (PV) cells and to extend the bandwidth of photon detectors[8,10–12]. Alternatively, the Schottky junction between metals and p-type semiconductors can be used to harvest hot holes generated by photons absorbed in the metal (Fig. 1b). However, the conversion efficiencies experimentally demonstrated to date have been extremely low.

Furthermore, it has been previously shown that the maximum *limiting* efficiency of the full solar spectrum harvesting and conversion via IPE from noble metals is restricted by the available electron density of states (e-DOS), because their e-DOS favors creation of large population of hot electrons with energies lower than the Schottky barrier height[13]. Prior theoretical estimate for a model metal with parabolic electron bands found the overall conversion efficiency limit at about 7%. In addition to the limits imposed on the harvesting of high-energy electrons by the available filled and empty electron energy levels, the high dark thermionic current through the Schottky junction at room temperature prevents the opportunity of lowering the energy barrier to increase the forward current and energy conversion efficiency.

Here, we estimate the *limiting* efficiency of the solar energy harvesting via hot electron photo-injection by using the realistic Au electron band-structure calculated via the first-principles method. Our data shows that the limiting efficiency reduces to a mere 3.6%, indicating the need to develop new photon-to-hot-carrier energy conversion schemes based on synergistic engineering of both photon and electron DOS in the broad energy range[14,15]. We further discuss possible ways to increase this limit, which include hot holes harvesting, optical concentration, and partial-spectrum conversion.

*sborisk@mit.edu; http://www.mit.edu/~sborisk/

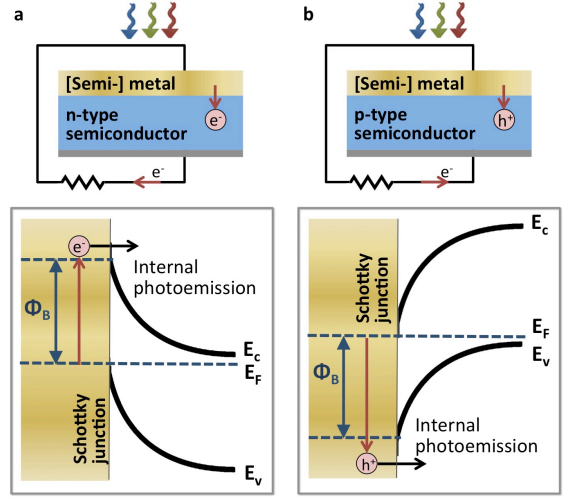

Figure 1. Schematics of the photon-to-electrical current energy conversion schemes using a metal photon absorber and the Schottky barrier as the hot-carrier filter. (a) Hot electron photoemission process across the junction between a metal and an n-type semiconductor. (a) Hot hole photoemission process across the junction between a metal and a p-type semiconductor.

## 2. FULL-SPECTRUM CONVERSION VIA INTERNAL PHOTOEMISSION THROUGH Au-TO-N-TYPE-SEMICONDUCTOR SCHOTTKY JUNCTION

Photon energy conversion efficiency of any solar harvesting device is affected by many contributing factors. In the case of the devices shown in Fig. 1, these factors include: efficiency of photon absorption, the probability of hot charge carriers to reach the materials junction before they thermalize, momentum matching of the hot carriers to make possible their transport through the barrier, etc. However, even assuming complete solar absorption, ballistic carriers transport to the junction and ideal momentum matching, the number of energy states available for electron transitions imposes the strongest limitation of the hot electron generation efficiency. As the photon-induced generation of the initial non-equilibrium hot charge carrier population is the first step in the energy conversion process, low efficiency of this step is almost impossible to overcome by optimizing subsequent steps.

A simple estimation of the limiting energy conversion efficiency of the device shown in Fig. 1a as a function of the absorber electron band-structure can be performed using the concept of electron distribution joint density of state (EDJDOS)[13]:

$$D(E, \hbar\omega) = \rho(E - \hbar\omega) \cdot f(E - \hbar\omega) \cdot \rho(E) \cdot (1 - f(E)). \qquad (1)$$

EDJDOS quantifies the number of initial $\left(\rho(E - \hbar\omega)\right)$ and final $\left(\rho(E)\right)$ electronic states in the material that are available for direct transitions given the absorbed photon energy. Here, $f(E)$ is the Fermi-Dirac distribution function, which defines the occupancy probability of an available energy level at a given temperature T.

In Figs. 2a,b, we compare e-DOS (blue) and EDJDOS (orange) in an ideal metal with parabolic energy levels (e-DOS~$E^{1/2}$)[13] with their counterparts in a realistic gold, calculated by using the Density Functional Theory (QUANTUM ESPRESSO simulation package[16]). Ideal metal is assumed to have a Fermi level at 5.5 eV from the conduction band bottom. The e-DOS calculation for a realistic case of bulk Au[17] was performed with the generalized gradient approximations for the exchange-correlation energy[18] and the projector augmented wave method[19]. Electron band structures were first obtained using Gaussian broadening method, and then interpolated onto a much finer mesh with the tetrahedra method to calculate the e-DOS[20]. In Figs. 2a,b, EDJDOS was calculated for the photon energy at the peak of the solar radiation ($\hbar\omega \sim 2eV$), and the initial electron states occupancy (blue shaded area) was calculated at T=300K.

As can be seen in Fig. 2a,b, a larger portion of hot electrons occupying higher-energy final states is created in the ideal metal rather than in gold (orange shaded area). Assuming that only hot electrons with energy higher than the Schottky barrier height can be internally emitted into the semiconductor with a non-negligible probability, we conclude that the energy distribution of the hot electron population in gold is extremely unfavorable for achieving high IPE efficiency.

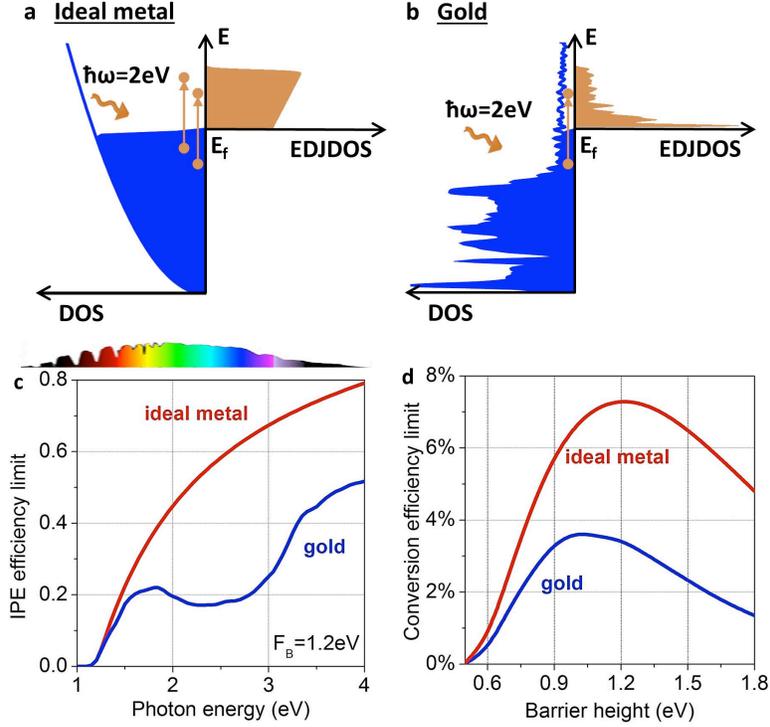

Figure 2. Comparison of the electron population of available energy states in the dark at T=300K (blue fill) and after the photo-excitation by photons with the energy $\hbar\omega = 2eV$ (orange fill) for (a) ideal metal with parabolic energy levels, (b) realistic bulk Au. The Au electron DOS was calculated by using the first-principles method. (c) Limiting IPE efficiency as a function of the absorbed photon energy and (d) the overall conversion efficiency under AM1.5D solar spectrum illumination as a function of the Schottky barrier height (red: ideal metal, blue: gold).

The IPE efficiency for both the ideal metal and Au is plotted in Fig. 2c as a function of the absorbed photon energy ($\hbar\omega$). It is calculated as a ratio of the population of the hot electrons with sufficient energy to be emitted over a Schottky barrier over the total photo-excited electron population:

$$\eta_{IPE}(\hbar\omega, \Phi_B) = \int_{\Phi_B}^{\hbar\omega} D(E, \hbar\omega) dE \cdot \left( \int_0^{\hbar\omega} D(E, \hbar\omega) dE \right)^{-1}. \qquad (2)$$

Figure 2c shows the IPE efficiency (i.e., the proportion of hot electrons with sufficient energy to be emitted over a Schottky barrier) for the case of the barrier height $\Phi_B = 1.2eV$. It can be seen that IPE efficiency is negligibly small for low-energy photons ($\hbar\omega \sim \Phi_B$), and grows with the increased photon energy. While the IPE efficiency grows monotonically in the case of the ideal metal, highly structured e-DOS in Au leads to the non-monotonic behavior of the IPE efficiency as a function of the absorbed photon energy. The parabolic approximation does not account for the interband transitions from the d-band energy levels at ~ 2.4 eV below the Fermi energy $E_F$. The localized nature of the d states leads to deviations from the free-electron parabolic trends for electron-to-electron scattering. However, these transitions dominate the EDJDOS of Au (Fig. 2b), which has a negative effect on the IPE efficiency (Fig. 2c).

To estimate the limiting efficiency, we assume that hot electrons travel ballistically to the junction and are subsequently emitted into the semiconductor to generate a photocurrent. This assumption can reasonably be applied in the cases when the thickness of the absorber is significantly thinner than the mean free path of the hot charge carrier, which for Au is about 40nm[21,22]. We also assume perfect momentum matching, which once again is justified for an Au absorber with characteristic dimensions much smaller that the carrier mean free path, as multiple reflections from the material interface increase the hot carrier emission probability.

The IPE efficiency $\eta_{IPE}(\hbar\omega, \Phi_B)$ can then be used to calculate the limiting value for the short circuit current generated by the hot electrons that have enough energy to be injected above the Schottky barrier into the semiconductor[13]:

$$J_{sc} = q \cdot \int_{solar} \eta_{IPE}(\hbar\omega, \Phi_B) \cdot \eta_{abs}(\hbar\omega) \cdot \varphi(\hbar\omega) d\omega. \quad (3)$$

Here, $q$ is the electron charge, $\varphi(\hbar\omega)$ is the incoming photon flux. To estimate the limiting efficiency value, we assume the ideal metal absorber: $\eta_{abs}(\hbar\omega) = 1$. The total maximum photo-generated current that can be delivered to the external load is reduced by the reverse current due to thermionic emission over the Schottky barrier, and is defined trough the standard Schottky diode equation as follows:

$$J = J_{sc} - A_R \cdot T^2 \cdot \exp(-\Phi_B/kT) \cdot (\exp(V/k_b T) - 1). \quad (4)$$

The thermionic emission reverse current scales as the second power with temperature, and depends on the specifics of the materials interface through the modified Richardson constant $A_R$. In the following calculations, we use the Richardson constant value previously reported for TiO$_2$ ($A_R = 6.71 \times 10^6 \, \text{Am}^{-2}\text{K}^{-2}$)[23], which is a common choice for semiconductor used in IPE devices. $V$ is the applied voltage, and $k_b$ is the Boltzmann constant

The overall limiting efficiency of the IPE converter is defined as the ratio of the maximum electrical power delivered to the load to the total power of the incoming sunlight ($I_{in} = \int_{solar} \varphi(\hbar\omega) d\omega = C \cdot \int_{solar} I_{AM1.5D}(\hbar\omega) d\omega$, $C$ being the solar concentration and $I_{AM1.5D}(\hbar\omega)$ - the tabulated AM1.5D terrestrial solar spectrum):

$$\eta = \max(J \cdot V)/I_{in}. \quad (5)$$

The maximum power point $(J_m, V_m)$ can be found by solving $d(J \cdot V)/dV = 0$.

Overall conversion efficiency for the incoming AM1.5D solar spectrum for Au and ideal metal is plotted in Fig. 2d as a function of the Schottky barrier height. These results show that the already low limiting efficiency value previously reported for the ideal metal[13] is reduced even further if the realistic electron density of states of Au is taken into account.

### 3. FULL AND PARTIAL SPECTRUM CONVERSION VIA INTERNAL PHOTOEMISSION THROUGH Au-TO-P-TYPE-SEMICONDUCTOR SCHOTTKY JUNCTION

The data shown in Fig. 3 illustrate that the full-spectrum energy converters based on the internal photo-emission of hot electrons through the Schottky junction between Au and an n-type semiconductor are severely limited by the available electron density of states in Au, and indicate that alternative strategies to improving the device efficiency should be pursued. One such strategy is based on the utilizing the process of the internal photo-injection of *hot holes* from Au to the adjacent p-type semiconductor, as schematically shown in Fig. 1b.

The calculated population of the photo-generated hot holes in Au indeed contains higher number of holes with the energies high enough to allow their emission across the Schottky barrier (see Fig. 1b and Fig. 3a). This translates into the higher limiting IPE efficiency and the overall conversion efficiency estimates for the Au solar converter based on the internal photo-emission of hot holes. By repeating the calculations outlined above for the device based on the hot hole collection, we indeed arrive at the higher limiting efficiency estimate, which raises above 10% (Fig. 3b, blue solid line).

Another commonly used approach to boosting the conversion efficiency is based on using the light concentration. Indeed, our calculations predict that for the illumination at 100 suns, the limiting efficiency maximum raises even more, reaching ~13% (Fig. 3b, blue dashed line).

Sorting photons by their energies and processing them separately typically helps to increase the light-to-work conversion efficiency in any solar energy conversion device[6,15,24]. Our calculations predict that spectral splitting schemes can also boost efficiency of hot carrier converters. This concept is illustrated in Fig. 3b for the case of the hot-hole IPE device. The data on Fig. 3b show that conversion of only high-energy and low-energy parts of the solar spectrum can be performed at the same limiting efficiency as the full-spectrum conversion (compare orange and blue lines). The center part of the spectrum can then be directed to a conventional photovoltaic cell to be converted at high efficiency (exceeding the cell conventional Shockley-Queisser[25] limit owing to elimination of the below-bandgap and high-energy electrons from the PV-converted partial spectrum[26]). Not surprisingly, the high-energy part of the solar spectrum alone (red lines in Fig. 3b) can be converted at much higher efficiency than the whole spectrum, while efficient conversion of the IR spectral portion alone is not feasible due to low energies of photo-generated hot holes (teal lines in Fig. 3b). These results indicate a promising way of using the solar converters based on the internal photoemission of hot carriers as a part of a hybrid system, especially at higher optical concentration.

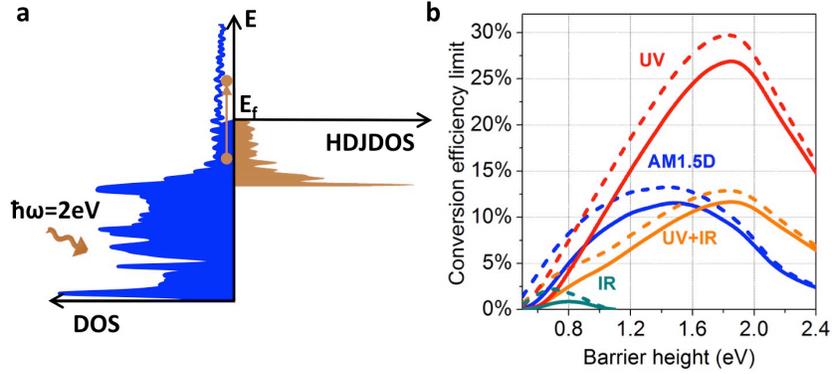

Figure 3. Electron population of available energy states in the dark at T=300K (blue fill) and the hot holes population after the photo-excitation by photons with the energy $\hbar\omega = 2eV$ (orange fill) in bulk Au. The Au electron DOS was calculated by using the first-principles method. (b) Limiting conversion efficiencies as a function of the Schottky barrier height under AM1.5D solar spectrum illumination as a function of the barrier height. These include: the efficiency of the full solar spectrum conversion (blue) and partial spectrum conversion of high-energy photons (red), low-energy infrared (IR) photons (teal), and both, high and low-energy parts of the spectrum (orange). Solid curves are for the solar concentration of one sun, dashed lines are for the concentration of 100 suns.

## 4. LIMITING RESPONSIVITIES OF Au HOT-CARRIER PHOTODETECTORS

Finally, we will estimate the efficiency of the Au hot-carrier devices for their use as photodetectors[8,10]. In this case, the power generation is not the final goal, and the photodetector responsivity limit can be calculated at the zero applied voltage, which results in the generation of the short circuit current defined by Eq. 3. The responsivity limits of hot-electron-based and hot-hole-based Au photodetectors are calculated as the ratios of the photo-generated currents per watt of incident radiant power. They are plotted in Fig. 4 as a function of the incoming photon wavelength and the Schottky barrier height.

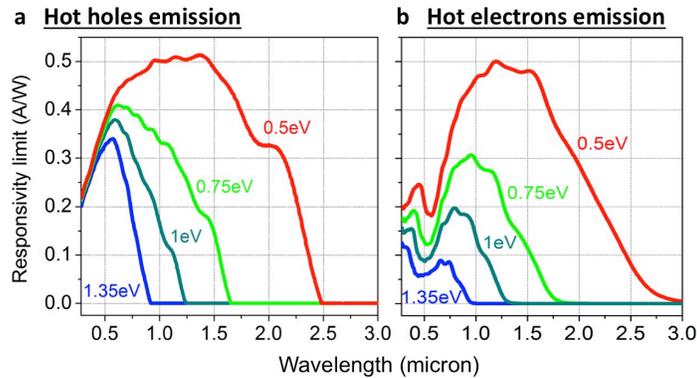

Figure 4. Limiting photodetection responsivity as a function of the absorbed photon energy for the Au detector based on the internal photo-injection of hot holes (a) and hot electrons (b). Individual curves correspond to the fixed values of the barrier height: $\Phi_B = 0.5eV$ (red), $\Phi_B = 0.75eV$ (green), $\Phi_B = 1eV$ (teal), and $\Phi_B = 1.35eV$ (blue).

Comparison of Fig. 4a and Fig. 4b reveals that the difference in the limiting responsivity values of the hot-electron and hot-hole devices is not as pronounced as in the case of the efficiency of solar energy conversion platforms. However, the hot-hole emission scheme still potentially offers higher limiting responsivity, especially for higher potential barriers at the materials interface.

## 5. DISCUSSION AND OUTLOOK

Our calculations of the limiting efficiencies of the solar energy converters based on light absorption in Au with the subsequent internal emission of the photo-generated hot carriers across the Schottky barrier revealed severe limitations to their potential. Even under the most favorable conditions, their performance is limited by the electron density of states in Au, and these limitations are even more stringent than those previously reported under the parabolic-bands approximation[13].

The efficiency of the Schottky-junction converter with Au absorber can theoretically be improved in the configuration when the hot holes are emitted into a semiconductor across the potential barrier. However, other issues may arise in this case, which may prevent the device from reaching the predicted value of the limiting efficiency. Although previous work reported comparable mean free path ranges for the hot electrons and holes in Au,[21] recent calculations of the total relaxation times of hot charge carriers in gold show that hot holes arising from $d$ states lose energy on a sub-5-femtosecond time scale[27], which makes hot $d$ holes challenging to extract before thermalization.

Some improvements to the limiting conversion efficiency may be achieved by the e-DOS modification via quantum confinement effects in low-dimensional Au absorbers[5,17]. Our preliminary calculations of the electron bandstructure indeed revealed some differences in the density of electron states in bulk Au and in thin Au films, with most significant changes observed in the higher-energy part of the energy spectrum. These effects may thus help to boost the efficiency of the high-energy photons conversion in a hybrid device with the spectral-splitting functionality. It should be noted here that hot carrier converters that make use of plasmonic nano-absorbers naturally benefit from high spectral as well as of high spatial selectivity, which translates into high optical concentration.

Overall, we can conclude that while the performance of the Au devices exploiting the internal photoemission of hot carriers is predicted to be low, some ways of boosting the performance have been identified. These include spectral splitting, optical concentration, and potentially quantum confinement effects. The above approaches should also be combined with the search for new cheap and abundant materials with electron characteristics more favorable for generating high-energy photo-excited charge carriers[14].

## ACKNOWLEDGEMENTS

This work has been supported by the US Department of Energy, Office of Science, and Office of Basic Energy via 'Solid State Solar-Thermal Energy Conversion Center (S3TEC)', Award No. DE-SC0001299/DE-FG02-09ER46577.